\documentclass[pra,aps,onecolumn,superscriptaddress,showpacs,showkeys,nofootinbib]{revtex4}
\usepackage{amsmath,amsthm,amssymb,graphicx,subfigure,xcolor}
\usepackage{mathrsfs}
\usepackage{graphics,float}
\usepackage{color}
\usepackage[unicode=true]{hyperref}
\usepackage{booktabs}
\usepackage{tabularx}
\usepackage{tabu}
\usepackage{multirow}
\hypersetup{
     colorlinks=true,         % false: boxed links; true: colored links
     linkcolor=blue,           % color of internal links
     citecolor=blue,            % color of links to bibliography
     urlcolor=blue,            % color of external links
 }
\newtheorem{theorem}{\hskip 1em Theorem}
\newtheorem*{theorem*}{Theorem}
\newtheorem{corollary}{\hskip 1em Corollary}
\newtheorem*{corollary*}{Corollary}

\newtheorem*{lemma*}{Lemma}

\newtheorem*{proposition*}{Proposition}
\theoremstyle{definition}

\newtheorem*{definition*}{Definition}
\theoremstyle{remark}

\newtheorem*{remark*}{Remark}
\begin{document}

\title{Methods of constructing superposition measures}

\author{Jialin Teng}
\author{Fengli Yan}
\email{flyan@hebtu.edu.cn}
\affiliation {College of Physics, Hebei Key Laboratory of Photophysics Research and Application, Hebei Normal University, Shijiazhuang 050024, China}
\author{Ting Gao}
\email{gaoting@hebtu.edu.cn}
\affiliation {School of Mathematics Science, Hebei Normal University, Shijiazhuang 050024, China}

\begin{abstract}
The resource theory of quantum superposition is an extension of the quantum coherent theory, in which linear independence relaxes the requirement of orthogonality. It can be used to quantify the nonclassical in superposition of finite number of optical coherent states.  Based on convex roof extended,  state transformation and  weight, we give three methods of constructing superposition measures of quantum states,  respectively. We also  generalize the superposition resource theory from two perspectives.
\end{abstract}

%\pacs{03.65.Ud}


\maketitle
%\tableofcontents{}

\section{INTRODUCTION}

Quantum information is a comprehensive  subject which combines quantum mechanics and information.  In quantum information, based on the basic principles of quantum mechanics, quantum states are used  to encode information and  the storage, processing and transmission of information are implemented by using quantum system.  Quantum communication \cite{1,2,3}, quantum cryptography \cite{4,5,6,7}, and quantum secret sharing \cite{8} are the main content of quantum information.

Quantum entanglement, first proposed by Einstein, Podolsky and Rosen \cite{9} in 1935, is a quantum correlation which can be regarded as a kind of resource. Because of the great success of entanglement theory \cite{21,22,Horodecki,18,23,24,25}, researchers naturally think whether the framework of entanglement theory can be applied to other fields.  In the quantum coherence,  {\AA}berg  proposed a  method of quantifying the superposition of orthogonal quantum states  in 2006 \cite{37}. In 2014, Baumgratz $et~ al$. established a rigorous framework of resource theory for quantifying quantum coherence \cite{38}. The frameworks of resource theory  in various fields have  been presented shortly \cite{35,17,36,26,27,28,29,30,31,32,33,34}.

Quantum resource theory provides a structural framework in quantum information  \cite{10,11,12,13,14,15,16,19,20}. Resource theory is determined by imposed constraints that may be due to fundamental conservation laws, such as super-selection rules and conservation of energy, or by practical difficulties in performing certain operations on the quantum states of system. The operations allowed by the constraints  are called free operation. In addition to free operation, there are another two elements in resource theory. One element is to measure the resources contained in the quantum states, such as entanglement and coherence. The other element is to define the so-called free state which does not contain resources. The free operation maps the free state to the free state.

In coherent resource theory, all the coherent resources used are considered with the condition that the base of the free states are orthogonal. In 2017, Theurer $et~ al$. proposed a generalized coherence theory called superposition resource theory, in which the requirement of orthogonality of the base of free states is extended to linear independence \cite{39}. This is of great significance for the extension of coherence theory. First, although linear independence relaxes the requirement for orthogonality, it still has a limitation requirement, which requires the basis to form a linearly independent set. So from a conceptual point of view, superposition resource theory helps clarify the difference between orthogonality and linear independence. Many of the results of coherence theory are special cases of results obtained in non-orthogonal environments.  This suggests that linear independence between free states, rather than orthogonality, is the main underlying cause of these results.  Second, superposition resource theory can quantify the nonclassical superposition of finite number of optical coherent states.  However, the coherent theoretical framework cannot be used in this case because the optical coherent states are not orthogonal. Therefore, superposition resource theory can be regarded as a new starting point and a  general resource theory.

In this paper we mainly study the methods of constructing superposition measures of  quantum states from different aspects. We organize this paper as follows.  In Sec.II, we give an overview of the superposition resource theory. In Sec.III, we present two methods for constructing superposition measure and obtain some new results. In Sec.IV, we provide the method of constructing superposition measure by weight. In Sec.V, we generalize the superposition resource theory. A brief summary is given in Sec.VI.

\section{overview of the superposition resource theory}\label{sec:review}

\subsection{Linearly independent  bases  and Gram matrix}\label{sec:review}

In superposition resource theory, one uses $\{|c_{i}\rangle\}_{i=1}^{d}$ to denote a set of linearly independent bases on Hilbert space, where the bases are not required to be orthogonal to each other. The Gram matrix is a useful tool for determining whether a given set of vectors is linearly independent \cite{40,41}. In the inner product space, given a set of finite vectors $\{v_{1},v_{2},\ldots,v_{m}\}$, the Gram matrix  is expressed as
\begin{equation}
\begin{aligned}
G= \begin{pmatrix} \langle v_{1}|v_{1}\rangle & \langle v_{1}|v_{2}\rangle & \cdots & \langle v_{1}|v_{m}\rangle \\ \langle v_{2}|v_{1}\rangle & \langle v_{2}|v_{2}\rangle & \cdots & \langle v_{2}|v_{m}\rangle \\ \vdots & \vdots & \ddots & \vdots \\ \langle v_{m}|v_{1}\rangle & \langle v_{2}|v_{m}\rangle & \ldots & \langle v_{m}|v_{m}\rangle \end{pmatrix}.
\end{aligned}
\end{equation}

A set of vectors is linearly independent if and only if the determinant of its  Gram matrix is positive \cite{42}.
For a quantum state set $\{|c_{1}\rangle,|c_{2}\rangle,\ldots,|c_{d}\rangle\}$, if $\langle c_{i}|c_{j}\rangle=\mu_{ij}$, its  Gram matrix becomes
\begin{equation}
\begin{aligned}
G= \begin{pmatrix} 1 & \mu_{12} & \cdots & \mu_{1d} \\ \mu_{12}^{*} & 1 & \cdots & \mu_{2d} \\ \vdots & \vdots & \ddots & \vdots \\ \mu_{1d}^{*} & \mu_{2d}^{*} & \ldots & 1 \end{pmatrix}.
\end{aligned}
\end{equation}
    If $\{|c_{1}\rangle,|c_{2}\rangle,\ldots,|c_{d}\rangle\}$ is linearly independent, then the determinant of the  Gram matrix is positive. In  special cases,  in which  $\langle c_{i}|c_{j}\rangle=\mu$ and $\mu$ be real numbers for arbitrary $i\neq j$,  the positive determinant of the Gram matrix  correspnds  $\mu\in(\frac{1}{1-d},1)$ for $d$-dimensional Hilber space.

\subsection{Free state and free operation}\label{sec:review}

The quantum state whose density matrix being $\rho=\sum\limits_{i=1}^{d}\rho_{i}|c_{i}\rangle\langle c_{i}|$ is called a free state, where $\{\rho_{i}\}$ is the probability distribution. The set of free states is denoted by $\mathcal{F}$. All quantum states except free states, are called resource states.

If the quantum operation $\Phi$ is trace-preserving and has the following form when it acts on quantum state $\rho$
\begin{equation}
\begin{aligned}
\Phi(\rho)=\sum\limits_{n}K_{n}\rho K_{n}^{\dag},
\end{aligned}
\end{equation}
then the quantum operation $\Phi$ is called free operation and the Kraus operator $K_{n}$ is called superposition-free, where $K_{n}=\sum_{k}c_{k,n}|c_{f_{n}(k)}\rangle\langle c_{k}^{\bot}|$, $c_{k,n}$ is a complex number, $f_{n}(k)$ is an index function, the quantum states $|c_{i}^{\bot}\rangle$ satisfy $\langle c_{i}^{\bot}|c_{j}\rangle=\xi_i\delta_{ij}$ for arbitrary $i,j=1,2,\cdots, d$, $\xi_i$ is a real number. The set of free operations is denoted by $\mathcal{FO}$.

\subsection{Superposition measure}\label{sec:review}

 A function $\mathcal{M}$ mapping all quantum states to the non-negative real numbers is called a superposition measure of quantum states if $\mathcal{M}$ satisfies the following conditions:

(S1) Faithful: $\mathcal{M}(\rho)=0$, if and only if $\rho\in\mathcal{F}$.

(S2) Monotonic under $\mathcal{FO}$: For any $\Lambda\in\mathcal{FO}$, $\mathcal{M}(\Lambda(\rho))\leq\mathcal{M}(\rho)$.

(S3) Monotonic under superposition-free selective measurements on average: $\sum_{n}p_{n}\mathcal{M}(\rho_{n})\leq\mathcal{M}(\rho)$, where $p_{n}={\rm Tr}(K_{n}\rho K_{n}^{\dagger})$, $\rho_{n}=\frac{K_{n}\rho K_{n}^{\dagger}}{p_{n}}$ for all $\{K_{n}\}$: $\sum\limits_{n}K_{n}^{\dag}K_{n}=\mathbb{I}$, $K_{n}\mathcal{F}K_{n}^{\dag}\subset\mathcal{F}$.

(S4) Convex: $\mathcal{M}(\sum\limits_{i}p_{i}\rho_{i})\leq\sum\limits_{i}p_{i}\mathcal{M}(\rho_{i})$, where the probability distribution $\{p_{i}\}$ satisfies $p_{i}\geq0$ and $\sum\limits_{i}p_{i}=1$.

A few valid superposition measures of quantum states  are listed in the following \cite{39}.

(1) The $l_{1}$ measure: $\mathcal{M}_{l_{1}}(\rho)=\sum\limits_{i\neq j}|\rho_{ij}|$,
for $\rho=\sum\limits_{i,j}\rho_{ij}|c_{i}\rangle\langle c_{j}|$.

(2) The relative entropy: $\mathcal{M}_{\rm rel.net}(\rho)=\min\limits_{\sigma\in\mathcal{F}}S(\rho\|\sigma)$,
where $S(\rho\|\sigma)={\rm Tr}(\rho\log\rho)-{\rm Tr}(\rho\log\sigma)$ denotes the quantum relative entropy.

(3) The rank-measure: $\mathcal{M}_{\rm rank}(|\psi\rangle)=\log r_{s}(|\psi\rangle)$, $\mathcal{M}_{\rm rank}(\rho)=\min\limits_{\{p_{n},|\varphi_{n}\rangle\}}\sum\limits_{n}p_{n}\mathcal{M}_{\rm rank}(|\varphi_{n}\rangle)$,
where $\{p_ {n},|\varphi_{n}\rangle\}$ is the ensemble of decomposition of $\rho$,  $r_{s}(|\psi\rangle)$ is the rank of the quantum state $|\psi\rangle$ with respect to the basis $\{c_{i}\}$.

(4) The robustness measure: $\mathcal{M}_{\rm R}(\rho)=\min\limits_{\tau~ \rm density ~matrix}\{s\geq 0|\frac{\rho+s\tau}{1+s}\in\mathcal{F}\}$.

\section{Superposition measure based on known function}

In this section we will provide two methods of constructing  superposition measure based on functions satisfying (S1) and (S3).

\subsection{Convex roof extended }\label{sec:review}

\begin{theorem} {\rm If a function $\mathcal{M}$ satisfies (S1) and (S3) for pure states, then the superposition measure can be obtained by the convex roof extended
\begin{equation}\label{7}
\begin{aligned}
\mathcal{M}^{\prime}(\rho)=\min\limits_{\{p_{n},|\varphi_{n}\rangle\}}\sum\limits_{n}p_{n}\mathcal{M}(|\varphi_{n}\rangle),
\end{aligned}
\end{equation}
where $\{p_ {n},|\varphi_{n}\rangle\}$ is the ensemble of decomposition of $\rho$.}
\end{theorem}

In order to simplify the expression, we use $\mathcal{M}(|\varphi\rangle)$ to represent $\mathcal{M}(|\varphi\rangle\langle\varphi|)$.
Here the function $\mathcal{M}$ satisfies (S1) for pure states means that for any pure state $|\psi\rangle$, there is $\mathcal{M}(|\psi\rangle)\geq0$, the equal sign holds if and only if $|\psi\rangle\in\mathcal{F}$, the set of free states; the function $\mathcal{M}$ satisfies (S3) for pure states means that the Kraus operator $\{K_{n}\}$ must satisfy
\begin{equation}
\begin{aligned}
\mathcal{M}(|\varphi\rangle)\geq\sum\limits_{n}{\rm Tr}(K_{n}|\varphi\rangle\langle\varphi|K_{n}^{\dag})
\mathcal{M}(\frac{K_{n}|\varphi\rangle\langle\varphi|K_{n}^{\dag}}{{\rm Tr}(K_{n}|\varphi\rangle\langle\varphi|K_{n}^{\dag})}).
\end{aligned}
\end{equation}

\textit{\rm \textbf{Proof.}} First we prove $\mathcal{M}^{\prime}(\rho)$ satisfies (S1). Because the function $\mathcal{M}$ satisfies (S1) for pure states, so
\begin{equation}
\begin{aligned}
\mathcal{M}^{\prime}(\rho)=\min\limits_{\{p_{n},|\varphi_{n}\rangle\}}\sum\limits_{n}p_{n}\mathcal{M}(|\varphi_{n}\rangle)\geq~0.
\end{aligned}
\end{equation}
If $\rho$ is a free state, then $\rho$ can be expressed as $\rho=\sum\limits_{i=1}^{d}p_{i}|c_{i}\rangle\langle c_{i}|$, so $\mathcal{M}^{\prime}(\rho)\leq\sum\limits_{i}p_{i}\mathcal{M}(|c_{i}\rangle)=0$, hence $\mathcal{M}^{\prime}(\rho)=0$.

If $\mathcal{M}^{\prime}(\rho)=0$, there is an ensemble decomposition of $\rho$ which is $\{p_{i}, |\psi_{i}\rangle\}$ makes
$\sum\limits_{i}p_{i}\mathcal{M}(|\psi_{i}\rangle)=0$. At this point, we have $\mathcal{M}(|\psi_{i}\rangle)=0$ for every $i$, so $\rho$ is free state.

Second we prove $\mathcal{M}^{\prime}(\rho)$ satisfies (S4). We want to prove that, for $\rho_{1}$ and $\rho_{2}$, there are $\mathcal{M}^{\prime}(p_{1}\rho_{1}+p_{2}\rho_{2})\leq p_{1}\mathcal{M}^{\prime}(\rho_{1})+p_{2}\mathcal{M}^{\prime}(\rho_{2})$, in which $p_{1}+p_{2}=1$. Assume that the set $\{p_{i}^{1}, |\psi_{i}^{1}\rangle\}$ is the optimal ensemble decomposition of $\rho_{1}$ and the set $\{p_{j}^{2}, |\psi_{j}^{2}\rangle\}$ is the optimal ensemble decomposition of $\rho_{2}$. That is to say $\mathcal{M}^{\prime}(\rho_{1})=\sum\limits_{i}p_{i}^{1}\mathcal{M}(|\varphi_{i}^{1}\rangle)$ and $\mathcal{M}^{\prime}(\rho_{2})=\sum\limits_{j}p_{j}^{2}\mathcal{M}(|\varphi_{j}^{2}\rangle)$.  Then one has
\begin{equation}
\begin{aligned}
p_{1}\mathcal{M}^{\prime}(\rho_{1})+p_{2}\mathcal{M}^{\prime}(\rho_{2})=\sum\limits_{i}p_{1}p_{i}^{1}\mathcal{M}(|\varphi_{i}^{1}\rangle)+\sum\limits_{j}p_{2}p_{j}^{2}\mathcal{M}(|\varphi_{j}^{2}\rangle).
\end{aligned}
\end{equation}
Because $p_{1}\rho_{1}+p_{2}\rho_{2}=\sum\limits_{i}p_{1}p_{i}^{1}|\varphi_{i}^{1}\rangle\langle\varphi_{i}^{1}|+\sum\limits_{j}p_{2}p_{j}^{2}|\varphi_{j}^{2}\rangle\langle\varphi_{j}^{2}|$, therefore
\begin{equation}
\begin{aligned}
\mathcal{M}^{\prime}(p_{1}\rho_{1}+p_{2}\rho_{2})&\leq\sum\limits_{i}p_{1}p_{i}^{1}\mathcal{M}(|\varphi_{i}^{1}\rangle)+\sum\limits_{j}p_{2}p_{j}^{2}\mathcal{M}(|\varphi_{j}^{2}\rangle)\\
&=p_{1}\mathcal{M}^{\prime}(\rho_{1})+p_{2}\mathcal{M}^{\prime}(\rho_{2}).
\end{aligned}
\end{equation}

After that  we prove $\mathcal{M}^{\prime}(\rho)$ satisfies (S3). Suppose the set $\{q_{m},|\varphi_{m}\rangle\}$ is the optimal ensemble decomposition of $\rho$. We have

\begin{equation}\label{a}
\begin{aligned}
\mathcal{M}^{\prime}(\rho)&=\sum\limits_{m}q_{m}\mathcal{M}(|\varphi_{m}\rangle)\\
&\geq\sum\limits_{m}q_{m}\sum\limits_{n}{\rm Tr}(K_{n}|\varphi_{m}\rangle\langle\varphi_{m}|K_{n}^{\dag})
\mathcal{M}(\frac{K_{n}|\varphi_{m}\rangle\langle\varphi_{m}|K_{n}^{\dag}}{{\rm Tr}(K_{n}|\varphi_{m}\rangle\langle\varphi_{m}|K_{n}^{\dag})})\\
&=\sum\limits_{n}{\rm Tr}(K_{n}\rho K_{n}^{\dag})
\sum\limits_{m}\frac{q_{m}{\rm Tr}(K_{n}|\varphi_{m}\rangle\langle\varphi_{m}|K_{n}^{\dag})}{{\rm Tr}(K_{n}\rho K_{n}^{\dag})}
\mathcal{M}(\frac{K_{n}|\varphi_{m}\rangle\langle\varphi_{m}|K_{n}^{\dag}}{{\rm Tr}(K_{n}|\varphi_{m}\rangle\langle\varphi_{m}|K_{n}^{\dag})})\\
&\geq\sum\limits_{n}{\rm Tr}(K_{n}\rho K_{n}^{\dag})\mathcal{M}^{\prime}(\frac{K_{n}\rho K_{n}^{\dag}}{{\rm Tr}(K_{n}\rho K_{n}^{\dag})}).
\end{aligned}
\end{equation}
The first  inequality in Eq.(\ref{a}) is based on  the fact that $\mathcal{M}$ satisfies (S3) for pure states. The second  inequality in Eq.(\ref{a}) comes from the fact that $\mathcal{M}^{\prime}(\rho)$ satisfies (S4).

Finally, we prove $\mathcal{M}^{\prime}$ satisfies (S2). Evidently
\begin{equation}\label{b}
\begin{aligned}
\mathcal{M}^{\prime}(\Lambda(\rho))\leq\sum\limits_{n}{\rm Tr}(K_{n}\rho K_{n}^{\dag})\mathcal{M}^{\prime}(\frac{K_{n}\rho K_{n}^{\dag}}{{\rm Tr}(K_{n}\rho K_{n}^{\dag})})\leq \mathcal{M}^{\prime}(\rho).
\end{aligned}
\end{equation}
The first  inequality in  Eq.(\ref{b}) takes the fact that $\mathcal{M}^{\prime}(\rho)$ satisfies (S4). The second  inequality in Eq.(\ref{b}) is the result of  the fact that $\mathcal{M}^{\prime}(\rho)$ satisfies (S3).

Thus we have proved the above theorem.

\begin{corollary}\label{1} {\rm $\mathcal{M}_{l_{1}}^{\prime}(\rho)=\min\limits_{\{p_{n},|\varphi_{n}\rangle\}}\sum\limits_{n}p_{n}\mathcal{M}_{l_{1}}(|\varphi_{n}\rangle)$ is a superposition measure, where $\{p_{n},|\varphi_{n}\rangle\}$ expresses an ensemble decomposition of $\rho$, $|\varphi_{n}\rangle=\sum\limits_{i}\varphi_{n,i}|c_{i}\rangle$, $\mathcal{M}_{l_{1}}(|\varphi_{n}\rangle)=\sum\limits_{i\neq j}|\varphi_{n,i}\varphi_{n,j}^{*}|$ stands for  the $l_{1}$ measure.}
\end{corollary}

\textit{\rm \textbf{Proof.}} Since $\mathcal{M}_{l_{1}}(\rho)$ is a superposition measure, hence $\mathcal{M}_{l_{1}}(\rho)$ satisfies (S1), (S2), (S3) and (S4) for any states. Thus $\mathcal{M}_{l_{1}}(|\varphi\rangle\langle\varphi|)$ satisfies (S1) and (S3) for all pure states. According to Theorem 1, Corollary 1 is true.

\begin{corollary}\label{1} {\rm For any state $\rho$, $\mathcal{M}_{l_{1}}^{\prime}(\rho)\geq\mathcal{M}_{l_{1}}(\rho)$.}
\end{corollary}

\textit{\rm \textbf{Proof.}} Suppose $\{q_{m},|\varphi_{m}\rangle\}$ is the optimal ensemble decomposition of $\rho$ for which $\mathcal{M}_{l_{1}}^{\prime}(\rho)$ is minimized.  Let $|\varphi_{m}\rangle=\sum\limits_{i}\varphi_{i}^{m}|c_{i}\rangle$, $\rho=\sum\limits_{i,j}\rho_{ij}|c_{i}\rangle\langle c_{j}|$. Then we have $\rho=\sum\limits_{m}q_{m}|\varphi_{m}\rangle\langle\varphi_{m}|=\sum\limits_{m}q_{m}\sum\limits_{i,j}\varphi_{i}^{m}{\varphi_{j}^{m}}^{*}|c_{i}\rangle\langle c_{j}|$. Therefore $\rho_{ij}=\sum\limits_{m}q_{m}\varphi_{i}^{m}{\varphi_{j}^{m}}^{*}$. One can easily derive
\begin{equation}
\begin{aligned}
\mathcal{M}_{l_{1}}(\rho)&=\sum\limits_{i\neq j}|\rho_{ij}|
=\sum\limits_{i\neq j}|\sum\limits_{m}q_{m}\varphi_{i}^{m}{\varphi_{j}^{m}}^{*}|\\
&\leq\sum\limits_{i\neq j}\sum\limits_{m}q_{m}|\varphi_{i}^{m}{\varphi_{j}^{m}}^{*}|
=\sum\limits_{m}q_{m}\sum\limits_{i\neq j}|\varphi_{i}^{m}{\varphi_{j}^{m}}^{*}|
=\sum\limits_{m}q_{m}\mathcal{M}_{l_{1}}(|\varphi_{m}\rangle\langle\varphi_{m}|)
=\mathcal{M}_{l_{1}}^{\prime}(\rho).
\end{aligned}
\end{equation}

\begin{corollary}\label{1} {\rm $\mathcal{M}_{\rm rel.net}^{\prime}(\rho)=\min\limits_{\{p_{n},|\varphi_{n}\rangle\}}\sum\limits_{n}p_{n}\mathcal{M}_{\rm rel.net}(|\varphi_{n}\rangle)$ is a superposition measure, where $\{p_{n},|\varphi_{n}\rangle\}$ is an ensemble decomposition of $\rho$, $\mathcal{M}_{\rm rel.net}(|\varphi_{n}\rangle)$ is the relative entropy.}
\end{corollary}

\textit{\rm \textbf{Proof.}} As $\mathcal{M}_{\rm rel.net}(\rho)$ is a superposition measure, therefore $\mathcal{M}_{\rm rel.net}(\rho)$ satisfies (S1), (S2), (S3) and (S4) for any states. Hence $\mathcal{M}_{\rm rel.net}(|\varphi\rangle\langle\varphi|)$ satisfies (S1) and (S3) for all pure states. By using Theorem 1, it is not difficult to confirm  Corollary 3.

\begin{theorem}{\rm Assume that some pure states form the set $\mathcal{S}$, and the convex combinations of states within $\mathcal{S}$ form the set $\mathcal{S}^{\prime}$. If a function $\mathcal{M}$ satisfies (S1) and (S3) for states in $\mathcal{S}$, then the superposition measure of all states in $\mathcal{S}^{\prime}$ can be obtained by the convex roof extended
\begin{equation}
\begin{aligned}
\mathcal{M}^{\prime}(\rho)=\min\limits_{\{p_{n},|\varphi_{n}\rangle\}}\sum\limits_{n}p_{n}\mathcal{M}(|\varphi_{n}\rangle),
\end{aligned}
\end{equation}
where $\{p_{n},|\varphi_{n}\rangle\}$ is the ensemble  decomposition of $\rho$, and meets $|\varphi_{n}\rangle\in\mathcal{S}$.}
\end{theorem}

The function $\mathcal{M}$ satisfies (S1) for states in $\mathcal{S}$ implies that for any $|\psi\rangle\in\mathcal{S}$, $\mathcal{M}(|\psi\rangle)\geq0$ is true if and only if $|\psi\rangle\in\mathcal{F}$ and $|\psi\rangle\in\mathcal{S}$; the function $\mathcal{M}$ satisfies (S3) for states in $\mathcal{S}$ means that the Kraus operator $\{K_{n}\}$ must satisfy $\frac{K_{n}|\psi\rangle\langle\psi|K_{n}^{\dag}}{{\rm Tr}(K_{n}|\psi\rangle\langle\psi|K_{n}^{\dag})}\in\mathcal{S}$ for any $|\psi\rangle\in\mathcal{S}$;  for $|\psi\rangle\in\mathcal{S}$ and $|\phi\rangle\in\mathcal{F}$, $\frac{K_{n}|\psi\rangle\langle\psi|K_{n}^{\dag}}{{\rm Tr}(K_{n}|\psi\rangle\langle\psi|K_{n}^{\dag})}\in\mathcal{S}$ and $\frac{K_{n}|\phi\rangle\langle\phi|K_{n}^{\dag}}{{\rm Tr}(K_{n}|\phi\rangle\langle\phi|K_{n}^{\dag})}\in\mathcal{F}$;  for $|\psi\rangle\in\mathcal{S}$,
\begin{equation}
\begin{aligned}
\mathcal{M}(|\psi\rangle)\geq\sum\limits_{n}{\rm Tr}(K_{n}|\psi\rangle\langle\psi|K_{n}^{\dag})
\mathcal{M}(\frac{K_{n}|\psi\rangle\langle\psi|K_{n}^{\dag}}{{\rm Tr}(K_{n}|\psi\rangle\langle\psi|K_{n}^{\dag})}).
\end{aligned}
\end{equation}

\textit{\rm \textbf{Proof.}}  First we prove $\mathcal{M}^{\prime}(\rho)$ satisfies (S1) for states in $\mathcal{S^{\prime}}$. It is equivalent to prove that $\mathcal{M}^{\prime}(\rho)=0$ if and only if $\rho\in\mathcal{F}$ and $\rho\in\mathcal{S^{\prime}}$.
For $|\varphi_{n}\rangle\in\mathcal{S}$, because the function $\mathcal{M}$ satisfies (S1) for states in $\mathcal{S}$, so
\begin{equation}
\begin{aligned}
\mathcal{M}^{\prime}(\rho)=\min\limits_{\{p_{n},|\varphi_{n}\rangle\}}\sum\limits_{n}p_{n}\mathcal{M}(|\varphi_{n}\rangle)\geq 0.
\end{aligned}
\end{equation}
If $\rho\in\mathcal{F}$, then $\rho$ can be expressed as $\rho=\sum\limits_{i=1}^{d}p_{i}|c_{i}\rangle\langle c_{i}|$, so
$\mathcal{M}^{\prime}(\rho)\leq\sum\limits_{i}p_{i}\mathcal{M}(|c_{i}\rangle)=0$, therefore $\mathcal{M}^{\prime}(\rho)=0$.

If $\mathcal{M}^{\prime}(\rho)=0$, there is an ensemble decomposition of $\rho$, $\{p_{i}, |\psi_{i}\rangle\}$ which makes
$\sum\limits_{i}p_{i}\mathcal{M}(|\psi_{i}\rangle)=0$. Since $\mathcal{M}(|\psi_{i}\rangle)\geq 0$, so we have $\mathcal{M}(|\psi_{i}\rangle)=0$ for every $i$, therefore $\rho\in\mathcal{F}$.

Second we prove $\mathcal{M}^{\prime}(\rho)$ satisfies (S4) for states in $\mathcal{S^{\prime}}$. For $\rho_{1}\in\mathcal{S^{\prime}}$ and $\rho_{2}\in\mathcal{S^{\prime}}$, evidently there is $\rho=p_{1}\rho_{1}+p_{2}\rho_{2}\in\mathcal{S^{\prime}}$, where $p_{1}+p_{2}=1$. We want to prove that, for $\rho_{1}\in\mathcal{S^{\prime}}$ and $\rho_{2}\in\mathcal{S^{\prime}}$, then $\mathcal{M}^{\prime}(p_{1}\rho_{1}+p_{2}\rho_{2})\leq p_{1}\mathcal{M}^{\prime}(\rho_{1})+p_{2}\mathcal{M}^{\prime}(\rho_{2})$, in which $p_{1}+p_{2}=1$. Assume that the set $\{p_{i}^{1}, |\psi_{i}^{1}\rangle\}$ is the optimal ensemble decomposition of $\rho_{1}$ and the set $\{p_{j}^{2}, |\psi_{j}^{2}\rangle\}$ is the optimal ensemble decomposition of $\rho_{2}$. It implies that $\mathcal{M}^{\prime}(\rho_{1})=\sum\limits_{i}p_{i}^{1}\mathcal{M}(|\varphi_{i}^{1}\rangle)$ and $\mathcal{M}^{\prime}(\rho_{2})=\sum\limits_{j}p_{j}^{2}\mathcal{M}(|\varphi_{j}^{2}\rangle)$.  Thus we have
\begin{equation}
\begin{aligned}
p_{1}\mathcal{M}^{\prime}(\rho_{1})+p_{2}\mathcal{M}^{\prime}(\rho_{2})=\sum\limits_{i}p_{1}p_{i}^{1}\mathcal{M}(|\varphi_{i}^{1}\rangle)+\sum\limits_{j}p_{2}p_{j}^{2}\mathcal{M}(|\varphi_{j}^{2}\rangle).
\end{aligned}
\end{equation}
Because $p_{1}\rho_{1}+p_{2}\rho_{2}=\sum\limits_{i}p_{1}p_{i}^{1}|\varphi_{i}^{1}\rangle\langle\varphi_{i}^{1}|+\sum\limits_{j}p_{2}p_{j}^{2}|\varphi_{j}^{2}\rangle\langle\varphi_{j}^{2}|$, so
\begin{equation}
\begin{aligned}
\mathcal{M}^{\prime}(p_{1}\rho_{1}+p_{2}\rho_{2})&\leq\sum\limits_{i}p_{1}p_{i}^{1}\mathcal{M}(|\varphi_{i}^{1}\rangle)+\sum\limits_{j}p_{2}p_{j}^{2}\mathcal{M}(|\varphi_{j}^{2}\rangle)\\
&=p_{1}\mathcal{M}^{\prime}(\rho_{1})+p_{2}\mathcal{M}^{\prime}(\rho_{2}).
\end{aligned}
\end{equation}

Next we prove $\mathcal{M}^{\prime}(\rho)$ satisfies (S3) for states in $\mathcal{S^{\prime}}$. Suppose set $\{q_{m},|\varphi_{m}\rangle\}$ is the optimal ensemble decomposition of $\rho$, therefore
\begin{equation}\label{c}
\begin{aligned}
\mathcal{M}^{\prime}(\rho)&=\sum\limits_{m}q_{m}\mathcal{M}(|\varphi_{m}\rangle)\\
&\geq\sum\limits_{m}q_{m}\sum\limits_{n}{\rm Tr}(K_{n}|\varphi_{m}\rangle\langle\varphi_{m}|K_{n}^{\dag})
\mathcal{M}(\frac{K_{n}|\varphi_{m}\rangle\langle\varphi_{m}|K_{n}^{\dag}}{{\rm Tr}(K_{n}|\varphi_{m}\rangle\langle\varphi_{m}|K_{n}^{\dag})})\\
&=\sum\limits_{n}{\rm Tr}(K_{n}\rho K_{n}^{\dag})
\sum\limits_{m}\frac{q_{m}{\rm Tr}(K_{n}|\varphi_{m}\rangle\langle\varphi_{m}|K_{n}^{\dag})}{{\rm Tr}(K_{n}\rho K_{n}^{\dag})}
\mathcal{M}(\frac{K_{n}|\varphi_{m}\rangle\langle\varphi_{m}|K_{n}^{\dag}}{{\rm Tr}(K_{n}|\varphi_{m}\rangle\langle\varphi_{m}|K_{n}^{\dag})})\\
&\geq\sum\limits_{n}{\rm Tr}(K_{n}\rho K_{n}^{\dag})\mathcal{M}^{\prime}(\frac{K_{n}\rho K_{n}^{\dag}}{{\rm Tr}(K_{n}\rho K_{n}^{\dag})}).
\end{aligned}
\end{equation}
The first  inequality in Eq.(\ref{c}) holds because  $\mathcal{M}$ satisfies (S3) for states in $\mathcal{S}$. The second inequality  in Eq.(\ref{c}) is true since  $\mathcal{M}^{\prime}(\rho)$ satisfies (S4) for states in $\mathcal{S^{\prime}}$.

Finally, we show $\mathcal{M}^{\prime}$ satisfies (S2) for states in $\mathcal{S^{\prime}}$. Evidently
\begin{equation}\label{d}
\begin{aligned}
\mathcal{M}^{\prime}(\Lambda(\rho))\leq\sum\limits_{n}{\rm Tr}(K_{n}\rho K_{n}^{\dag})\mathcal{M}^{\prime}(\frac{K_{n}\rho K_{n}^{\dag}}{{\rm Tr}(K_{n}\rho K_{n}^{\dag})})\leq \mathcal{M}^{\prime}(\rho).
\end{aligned}
\end{equation}
The first  inequality in Eq.(\ref{d}) comes from  the fact that $\mathcal{M}^{\prime}(\rho)$ satisfies (S4) for states in $\mathcal{S^{\prime}}$. The second inequality in Eq.(\ref{d}) holds since $\mathcal{M}^{\prime}(\rho)$ satisfies (S3) for states in $\mathcal{S^{\prime}}$.

The proof of Theorem 2 has been completed.

\subsection{State transformation method}\label{sec:review}

In this section we will give another method of constructing superposition measure called the state transformation method, which is stated as the following theorem.

\begin{theorem} {\rm For $i\neq j$, let $\langle c_{i}|c_{j}\rangle=\mu$ and $\mu$ be real numbers. If a function $\mathcal{M}$ satisfies (S1) and (S3) for pure states and $\mathcal{M}$ satisfies $\mathcal{M}(|\varphi_{0}\rangle)\leq\sum\limits_{i}p_{i}\mathcal{M}(|\psi_{i}\rangle)$, the superposition measure can be obtained by
\begin{equation}
\begin{aligned}
\Gamma(\rho)=\inf\limits_{|\phi\rangle\in R(\rho)}\mathcal{M}(|\phi\rangle).
\end{aligned}
\end{equation}
Here $R(\rho)$ is the set of all pure states that can be converted to $\rho$ by free operation, $|\varphi_{0}\rangle\in R(\rho)$,  $\{p_{i},|\psi_{i}\rangle\}$ is any ensemble decomposition of $\rho$ \cite{43}.}
\end{theorem}

\textit{\rm \textbf{Proof.}}  First we prove $\Gamma(\rho)$ satisfies (S1) for any state.

Obviously, $\Gamma(\rho)\geq 0$.  Evidently, an arbitrary  free state can be expressed as  $\rho'=\sum\limits_{i=1}^{d}\rho_{i}|c_{i}\rangle\langle c_{i}|$, where $\rho_i\geq 0$ and  $\sum\limits_{i=1}^{d}\rho_{i}=1$. Choose Kraus operators as
\begin{equation}
\begin{aligned}
K_{1}=\sqrt{\rho_{1}}(\frac {1}{\xi_1}|c_{1}\rangle\langle c_{1}^{\bot}|+
\frac {1}{\xi_2}|c_{2}\rangle\langle c_{2}^{\bot}|&+\cdots+\frac {1}{\xi_{d-1}}|c_{d-1}\rangle\langle c_{d-1}^{\bot}|+\frac {1}{\xi_d}|c_{d}\rangle\langle c_{d}^{\bot}|),\\
K_{2}=\sqrt{\rho_{2}}(\frac {1}{\xi_1}|c_{2}\rangle\langle c_{1}^{\bot}|+
\frac {1}{\xi_2}|c_{3}\rangle\langle c_{2}^{\bot}|&+\cdots+\frac {1}{\xi_{d-1}}|c_{d}\rangle\langle c_{d-1}^{\bot}|+\frac {1}{\xi_d}|c_{1}\rangle\langle c_{d}^{\bot}|),\\
K_{3}=\sqrt{\rho_{3}}(\frac {1}{\xi_1}|c_{3}\rangle\langle c_{1}^{\bot}|+
\frac {1}{\xi_2}|c_{4}\rangle\langle c_{2}^{\bot}|&+\cdots+\frac {1}{\xi_{d-1}}|c_{1}\rangle\langle c_{d-1}^{\bot}|+\frac {1}{\xi_d}|c_{2}\rangle\langle c_{d}^{\bot}|),\\
&\vdots\\
K_{d}=\sqrt{\rho_{d}}(\frac {1}{\xi_1}|c_{d}\rangle\langle c_{1}^{\bot}|+
\frac {1}{\xi_2}|c_{1}\rangle\langle c_{2}^{\bot}|&+\cdots+\frac {1}{\xi_{d-1}}|c_{d-2}\rangle\langle c_{d-1}^{\bot}|+\frac {1}{\xi_d}|c_{d-1}\rangle\langle c_{d}^{\bot}|),
\end{aligned}
\end{equation}
where $\xi_i=\langle c_i^{\bot}|c_i\rangle$. It is easy to prove that $\{K_1,K_2,\cdots, K_d\}$ is a free operation, $\sum\limits_{i=1}^dK_i^+K_i=I$ and  $\rho'=\sum\limits_{i=1}^{d}\rho_{i}|c_{i}\rangle\langle c_{i}|=\sum\limits_{i=1}^{d}K_{i}|c_{1}\rangle\langle c_{1}|K_{i}^{\dag}$. Thus we have
\begin{equation}
0\leq\Gamma(\rho')\leq\Gamma(|c_1\rangle)=0.
\end{equation}
It means $\Gamma(\rho')=0$.

Suppose that $\Gamma(\rho)=0$. According to the definition, there must be one pure state $|\phi\rangle$ and free operation $\Lambda$, such that $\mathcal{M}(|\phi\rangle) =0$ and $\rho=\Lambda(|\phi\rangle\langle\phi|)$. It implies that both $|\phi\rangle$ and $\rho$ are free states. Therefore, the  superposition measure satisfies (S1).

Next we prove $\Gamma(\rho)$ satisfies (S2).

A state undergoing two free operations can be regarded as undergoing one free operation. For $\rho_{0}=\Lambda(\rho)=\Lambda[\varepsilon(|\psi\rangle\langle\psi|)]$, where $\Lambda$ and $\varepsilon$ are free operations, $\rho_{0}$ and $\rho$ can be obtained from $|\psi\rangle$ by free operations.  If $|\psi\rangle$ is optimal for $\rho$, then $\Gamma(\rho)=\mathcal{M}(|\psi\rangle)$. Because $|\psi\rangle$ is not necessarily optimal for $\rho_{0}$, so $\Gamma(\rho_{0})\leq~\mathcal{M}(|\psi\rangle)$. Hence $\Gamma(\rho_{0})\leq\Gamma(\rho)$. Thus we have shown that $\Gamma(\rho)$ satisfies (S2).

Then we will prove $\Gamma(\rho)$ satisfies (S4). We show that $\Gamma(\rho)$ satisfies (S4) by proving that $\Gamma(\rho)=\mathcal{M}^{\prime}(\rho)=\min\limits_{\{p_{i},|\psi_{i}\rangle\}}\sum\limits_{i}p_{i}\mathcal{M}(|\psi_{i}\rangle)$.

 Firstly, we prove that for any superposition measure $\mathcal{C}$, if $\mathcal{M}(|\psi\rangle)=\mathcal{C}(|\psi\rangle)$ for all pure states, then $\Gamma(\rho)\geq \mathcal{C}(\rho)$ for any state $\rho$. Assume that  $|\psi\rangle$ is optimal for $\rho$, then $\Gamma(\rho)=\mathcal{M}(|\psi\rangle)$, and we know that $|\psi\rangle$ can be transformed into $\rho$ by free operations, then $\mathcal{C}(\rho)\leq~\mathcal{C}(|\psi\rangle)=\mathcal{M}(|\psi\rangle)$ is given by the monotonicity of $\mathcal{C}$.  So $\Gamma(\rho)\geq~\mathcal{C}(\rho)$.

Second we show that when $\mathcal{M}$ satisfies $\mathcal{M}(|\varphi_{0}\rangle)\leq\sum\limits_{i}p_{i}\mathcal{M}(|\psi_{i}\rangle)$, there is $\Gamma(\rho)=\mathcal{M}^{\prime}(\rho)$.

Obviously, $\mathcal{M}^{\prime}(\rho)=\min\limits_{\{p_{i},|\psi_{i}\rangle\}}\sum\limits_{i}p_{i}\mathcal{M}(|\psi_{i}\rangle)$ is one of superposition measure $\mathcal{C}$. So we have $\Gamma(\rho)\geq \mathcal{M}^{\prime}(\rho)$. By $\mathcal{M}(|\varphi_{0}\rangle)\leq\sum\limits_{i}p_{i}\mathcal{M}(|\psi_{i}\rangle)$, we know that $\Gamma(\rho)\leq\mathcal{M}(|\varphi_{0}\rangle)\leq\mathcal{M}^{\prime}(\rho)$. Therefore when $\mathcal{M}$ satisfies $\mathcal{M}(|\varphi_{0}\rangle)\leq\sum\limits_{i}p_{i}\mathcal{M}(|\psi_{i}\rangle)$, we have $\Gamma(\rho)=\mathcal{M}^{\prime}(\rho)$. Because $\mathcal{M}^{\prime}(\rho)$ satisfies (S4), thus $\Gamma(\rho)$ meets (S4).

Finally, we prove $\Gamma(\rho)$ satisfies (S3). Suppose $|\psi\rangle$ is converted to $\rho=\sum\limits_{i}t_{i}|\varphi_{i}\rangle\langle\varphi_{i}|$ by free operation, and then converted to $\{p_{l},\rho_{l}\}=\{t_{i}q_{il},|\phi_{il}\rangle\}$ by free operation. Suppose $|\psi\rangle$ is optimal for $\rho$, where $p_{l}=\sum\limits_{i}t_{i}q_{il}$, $\rho_{l}=\sum\limits_{l}\frac{t_{i}q_{il}}{p_{l}}|\phi_{il}\rangle\langle\phi_{il}|$, there is
\begin{equation}\label{e}
\begin{aligned}
\Gamma(\rho)&=\mathcal{M}(|\psi\rangle)
\geq\sum\limits_{i,l}t_{i}q_{il}\mathcal{M}(|\phi_{il}\rangle)
=\sum\limits_{l}p_{l}\sum\limits_{i}\frac{t_{i}q_{il}}{p_{l}}\mathcal{M}(|\phi_{il}\rangle)
=\sum\limits_{l}p_{l}\sum\limits_{i}\frac{t_{i}q_{il}}{p_{l}}\Gamma(|\phi_{il}\rangle)
\geq\sum\limits_{l}p_{l}\Gamma(\rho_{l}).
\end{aligned}
\end{equation}
The first  inequality in Eq.(\ref{e}) ia based on  the fact that $\mathcal{M}$ satisfies (S3) for pure states. Another  inequality in above equation  holds by the fact that $\Gamma(\rho)$ meets (S4).

Thus we have proved Theorem 3.

\textit{\em \textbf{Example 1.}} For qubit, let us consider a class of states $\rho(x)=\frac{1}{1+2\mu x}(\frac{1}{2}|c_{0}\rangle\langle c_{0}|+x|c_{0}\rangle\langle c_{1}|+x|c_{1}\rangle\langle c_{0}|+\frac{1}{2}|c_{1}\rangle\langle c_{1}|)$, where $-1<\mu=\langle c_{0}|c_{1}\rangle<1$ and $-\frac{1}{2}\leq x\leq\frac{1}{2}$. Choose $\mathcal{M}_{l_{1}}(\rho)$ as the function that satisfies (S1) and (S3) for pure states. The quantum state $|\varphi_{0}\rangle$  for the state $\rho(x)$  will be given as follows \cite{44}.

 Apparently,  quantum state $\rho(x)$ can be decomposed as
\begin{equation}
\begin{aligned}
\rho(x)=p_{1}(x,\alpha)|\psi_{1}(x,\alpha)\rangle\langle \psi_{1}(x,\alpha)|+p_{2}(x,\alpha)|\psi_{2}(x,\alpha)\rangle\langle \psi_{2}(x,\alpha)|,
\end{aligned}
\end{equation}
and each $\alpha$ corresponds a decomposition of $\rho(x)$, where
\begin{equation}
\begin{aligned}
&\sqrt{p_{1}(x,\alpha)}|\psi_{1}(x,\alpha)\rangle=\cos\alpha\sqrt{\lambda_{1}(x)}|c_{+}\rangle+\sin\alpha\sqrt{\lambda_{2}(x)}|c_{-}\rangle,\\
&\sqrt{p_{2}(x,\alpha)}|\psi_{2}(x,\alpha)\rangle=-\sin\alpha\sqrt{\lambda_{1}(x)}|c_{+}\rangle+\cos\alpha\sqrt{\lambda_{2}(x)}|c_{-}\rangle,\\
&p_{1}(x,\alpha)=\cos^{2}\alpha\lambda_{1}(x)+\sin^{2}\alpha\lambda_{2}(x),\\
&p_{2}(x,\alpha)=\sin^{2}\alpha\lambda_{1}(x)+\cos^{2}\alpha\lambda_{2}(x),\\
&\lambda_{1}(x)=\frac{(1+\mu)(1+2x)}{2+4\mu x},\\
&\lambda_{2}(x)=\frac{(1-\mu)(1-2x)}{2+4\mu x},\\
&|c_{+}\rangle=\frac{1}{\sqrt{2\mu+2}}(|c_{0}\rangle+|c_{1}\rangle),~~|c_{-}\rangle=\frac{1}{\sqrt{2-2\mu}}(|c_{0}\rangle-|c_{1}\rangle).\\
\end{aligned}
\end{equation}
So we have
\begin{equation}\label{f}
\begin{aligned}
p_{1}(x,\alpha)\mathcal{M}_{l_{1}}(|\psi_{1}(x,\alpha)\rangle)+p_{2}(x,\alpha)\mathcal{M}_{l_{1}}(|\psi_{2}(x,\alpha)\rangle)
=\frac{1}{2+4\mu x}(|2x+\cos2\alpha|+|2x-\cos2\alpha|)
\geq\frac{|2x|}{1+2\mu x}.
\end{aligned}
\end{equation}
Obviously $\sum\limits_{i=1}^2p_{i}(x,\alpha)\mathcal{M}_{l_{1}}(|\psi_{i}(x,\alpha)\rangle)$ is a function of $\alpha$. When $\alpha=\frac{\pi}{4}$, the equal sign holds in Eq.(\ref{f}). Therefore, we find that the optical ensemble decomposition of $\rho$ which  minimizes $\sum\limits_{i=1}^2p_{i}(x,\alpha)\mathcal{M}_{l_{1}}(|\psi_{i}(x,\alpha)\rangle)$, corresponds to   $\alpha=\frac{\pi}{4}$ and yields
\begin{equation}
\begin{aligned}
p_{1}(x,\frac{\pi}{4})&=p_{2}(x,\frac{\pi}{4})=\frac{1}{2}(\lambda_{1}(x)+\lambda_{2}(x))=\frac{1}{2},\\
|\psi_{1}(x,\frac{\pi}{4})\rangle&=\frac{\frac{\sqrt{\lambda_{1}(x)}}{\sqrt{4\mu+4}}+\frac{\sqrt{\lambda_{2}(x)}}{\sqrt{4-4\mu}}}{\sqrt{\frac{1}{2}(\lambda_{1}(x)+\lambda_{2}(x))}}|c_{0}\rangle
+\frac{\frac{\sqrt{\lambda_{1}(x)}}{\sqrt{4\mu+4}}-\frac{\sqrt{\lambda_{2}(x)}}{\sqrt{4-4\mu}}}{\sqrt{\frac{1}{2}(\lambda_{1}(x)+\lambda_{2}(x))}}|c_{1}\rangle,\\
|\psi_{2}(x,\frac{\pi}{4})\rangle&=\frac{-\frac{\sqrt{\lambda_{1}(x)}}{\sqrt{4\mu+4}}+\frac{\sqrt{\lambda_{2}(x)}}{\sqrt{4-4\mu}}}{\sqrt{\frac{1}{2}(\lambda_{1}(x)+\lambda_{2}(x))}}|c_{0}\rangle
+\frac{-\frac{\sqrt{\lambda_{1}(x)}}{\sqrt{4\mu+4}}-\frac{\sqrt{\lambda_{2}(x)}}{\sqrt{4-4\mu}}}{\sqrt{\frac{1}{2}(\lambda_{1}(x)+\lambda_{2}(x))}}|c_{1}\rangle.\\
\end{aligned}
\end{equation}

Thus if we choose $|\varphi_{0}\rangle=|\psi_{1}(x,\frac{\pi}{4})\rangle$, then we have that $|\varphi_{0}\rangle$  satisfies
\begin{equation}
\begin{aligned}
\Gamma(\rho(x))=\mathcal{M}_{l_{1}}(|\varphi_{0}\rangle)\leq\sum\limits_{i=1}^2p_{i}(x,\alpha)\mathcal{M}_{l_{1}}(|\psi_{i}(x,\alpha)\rangle).
\end{aligned}
\end{equation}

We choose the free operation $\{K_{1}=\sqrt{\frac {1}{2}}(\frac {|c_{0}\rangle\langle c_{0}^{\bot}|}{\langle c_{0}^{\bot}|c_{0}\rangle}+
\frac {|c_{1}\rangle\langle c_{1}^{\bot}|}{\langle c_{1}^{\bot}|c_{1}\rangle}), K_{2}=\sqrt{\frac {1}{2}}(\frac {|c_{1}\rangle\langle c_{0}^{\bot}|}{\langle c_{0}^{\bot}|c_{0}\rangle}+
\frac{|c_{0}\rangle\langle c_{1}^{\bot}|}{\langle c_{1}^{\bot}|c_{1}\rangle})\}$. It is easy to deduce that
\begin{equation}
\rho(x)=\sum_{i=1}^2K_i|\varphi_0\rangle\langle \varphi_0|K_i^+.
\end{equation}

Therefore, the $|\varphi_0\rangle$ which meets the requirements has been found.

\section{Superposition measure based on weight}\label{sec:review}

In superposition resource theory, we can divide each state into a free part and a resource part \cite{45}, one can  construct the following superposition measure based on weight. For quantum state $\rho$, we define
\begin{equation}
\begin{aligned}
\mathcal{M}_{w}(\rho)=\min\limits_{\tau~{\rm density ~matrix}}\{1-\lambda|\rho=\lambda\delta+(1-\lambda)\tau,\delta\in \mathcal{F}\}.
\end{aligned}
\end{equation}
Evidently, $0\leq\mathcal{M}_{w}(\rho)\leq 1$. Then we arrive at the following result.

\begin{theorem}\label{2}
{\rm $\mathcal{M}_{w}(\rho)$  is a superposition measure.}
\end{theorem}

\textit{\rm \textbf{Proof.}} Clearly,   $\mathcal{M}_{w}(\rho)$ satisfies (S1).

Next we prove $\mathcal{M}_{w}(\rho)$ meets (S4). It implies that we want to prove $\mathcal{M}_{w}(\rho)\leq p\mathcal{M}_{w}(\rho_{1})+(1-p)\mathcal{M}_{w}(\rho_{2})$, where $\rho=p\rho_{1}+(1-p)\rho_{2}$. Suppose $\widetilde{\delta_{1}}$ and $\widetilde{\tau_{1}}$ are the optimal states for $\rho_{1}$ to obtain $\mathcal{M}_{w}(\rho_{1})$, and suppose $\widetilde{\delta_{2}}$ and $\widetilde{\tau_{2}}$ are the optimal states for $\rho_{2}$ to obtain $\mathcal{M}_{w}(\rho_{2})$, where $\widetilde{\delta_{1}}$ and $\widetilde{\delta_{2}}$ are free states, $\widetilde{\tau_{1}}$ and $\widetilde{\tau_{2}}$ are resource states.  So we have $\rho_{1}=[1-\mathcal{M}_{w}(\rho_{1})]\widetilde{\delta_{1}}+\mathcal{M}_{w}(\rho_{1})\widetilde{\tau_{1}}$, $\rho_{2}=[1-\mathcal{M}_{w}(\rho_{2})]\widetilde{\delta_{2}}+\mathcal{M}_{w}(\rho_{2})\widetilde{\tau_{2}}$. Without loss of generality, we assume $\mathcal{M}_{w}(\rho_{1})\neq 0$ and $\mathcal{M}_{w}(\rho_{2})\neq 0$. Hence

\begin{equation}\label{3}
\begin{array}{ll}
\rho&=p\rho_{1}+(1-p)\rho_{2}\\
&=p([1-\mathcal{M}_{w}(\rho_{1})]\widetilde{\delta_{1}}+\mathcal{M}_{w}(\rho_{1})\widetilde{\tau_{1}})+(1-p)([1-\mathcal{M}_{w}(\rho_{2})]\widetilde{\delta_{2}}+\mathcal{M}_{w}(\rho_{2})\widetilde{\tau_{2}})\\
&=(1-M)\frac{p[1-\mathcal{M}_{w}(\rho_{1})]\widetilde{\delta_{1}}+(1-p)[1-\mathcal{M}_{w}(\rho_{2})]\widetilde{\delta_{2}}}{1-M}+M\frac{p\mathcal{M}_{w}(\rho_{1})\widetilde{\tau_{1}}+(1-p)\mathcal{M}_{w}(\rho_{2})\widetilde{\tau_{2}}}{M},
\end{array}
\end{equation}\\
where $ M =p\mathcal{M}_{w}(\rho_{1})+(1-p)\mathcal{M}_{w}(\rho_2).$

Note that $\widetilde{\delta_{1}}$ and $\widetilde{\delta_{2}}$ are  free states, ${\rm Tr}(p[1-\mathcal{M}_{w}(\rho_{1})]\widetilde{\delta_{1}}+(1-p)[1-\mathcal{M}_{w}(\rho_{2})]\widetilde{\delta_{2}})=1-(p\mathcal{M}_{w}(\rho_{1})+(1-p)\mathcal{M}_{w}(\rho_{2}))=1-M$
, so $\frac{p[1-\mathcal{M}_{w}(\rho_{1})]\widetilde{\delta_{1}}+(1-p)[1-\mathcal{M}_{w}(\rho_{2})]\widetilde{\delta_{2}}}{1-M}$ is a free state, and $\frac{p\mathcal{M}_{w}(\rho_{1})\widetilde{\tau_{1}}+(1-p)\mathcal{M}_{w}(\rho_{2})\widetilde{\tau_{2}}}{M}$ is a resource state.  Then $(1-M)\frac{p[1-\mathcal{M}_{w}(\rho_{1})]\widetilde{\delta_{1}}+(1-p)[1-\mathcal{M}_{w}(\rho_{2})]\widetilde{\delta_{2}}}{1-M}+M\frac{p\mathcal{M}_{w}(\rho_{1})\widetilde{\tau_{1}}
+(1-p)\mathcal{M}_{w}(\rho_{2})\widetilde{\tau_{2}}}{M}$ is an ensemble decomposition of $\rho$.  According to the definition of $\mathcal{M}_{w}(\rho)$, we know that $\mathcal{M}_{w}(\rho)\leq M$. It means that  $\mathcal{M}_{w}(\rho)$ satisfies (S4).

Later on we will show $\mathcal{M}_{w}(\rho)$ satisfies (S3). That is  one  wants to prove  $\sum\limits_{n}{\rm Tr}(K_{n}\rho K_{n}^{\dag})\mathcal{M}_{w}(\frac{K_{n}\rho K_{n}^{\dag}}{{\rm Tr}[K_{n}\rho K_{n}^{\dag}]})\leq \mathcal{M}_{w}(\rho)$, where $K_n$ is superposition-free. Suppose $\widetilde{\delta}$ and $\widetilde{\tau}$ are the optimal states for $\rho$ to minimize $\mathcal{M}_{w}(\rho)$, where $\widetilde{\delta}$ is a free state and $\widetilde{\tau}$ is a resource state. Then we obtain
$\rho=[1-\mathcal{M}_{w}(\rho)]\widetilde{\delta}+\mathcal{M}_{w}(\rho)\widetilde{\tau}$. Furthermore
\begin{equation}
\begin{aligned}
K_{n}\rho K_{n}^{\dag}=[1-\mathcal{M}_{w}(\rho)]K_{n}\widetilde{\delta} K_{n}^{\dag}+\mathcal{M}_{w}(\rho)K_{n}\widetilde{\tau} K_{n}^{\dag}.
\end{aligned}
\end{equation}
Therefore, we get
\begin{equation}
\begin{aligned}
\frac{K_{n}\rho K_{n}^{\dag}}{{\rm Tr}(K_{n}\rho K_{n}^{\dag})}=\frac{[1-\mathcal{M}_{w}(\rho)]K_{n}\widetilde{\delta}K_{n}^{\dag}}{{\rm Tr}(K_{n}\rho K_{n}^{\dag})}+\frac{\mathcal{M}_{w}(\rho)K_{n}\widetilde{\tau}K_{n}^{\dag}}{{\rm Tr}(K_{n}\rho K_{n}^{\dag})}.
\end{aligned}
\end{equation}
So
\begin{equation}\label{q}
\begin{array}{ll}
\mathcal{M}_{w}(\frac{K_{n}\rho K_{n}^{\dag}}{{\rm Tr}(K_{n}\rho K_{n}^{\dag})})&=\mathcal{M}_{w}(\frac{[1-\mathcal{M}_{w}(\rho)]K_{n}\widetilde{\delta}K_{n}^{\dag}}{{\rm Tr}(K_{n}\rho K_{n}^{\dag})}+\frac{\mathcal{M}_{w}(\rho)K_{n}\widetilde{\tau}K_{n}^{\dag}}{{\rm Tr}(K_{n}\rho K_{n}^{\dag})})\\
&\leq\mathcal{M}_{w}({\rm Tr}(\frac{[1-\mathcal{M}_{w}(\rho)]K_{n}\widetilde{\delta}K_{n}^{\dag}}{{\rm Tr}(K_{n}\rho K_{n}^{\dag})})
\frac{[1-\mathcal{M}_{w}(\rho)]K_{n}\widetilde{\delta}K_{n}^{\dag}}{{\rm Tr}(K_{n}\rho K_{n}^{\dag}){\rm Tr}(\frac{[1-\mathcal{M}_{w}(\rho)]K_{n}\widetilde{\delta}K_{n}^{\dag}}{{\rm Tr}(K_{n}\rho K_{n}^{\dag})})})\\
&\quad+\mathcal{M}_{w}({\rm Tr}(\frac{\mathcal{M}_{w}(\rho)K_{n}\widetilde{\tau}K_{n}^{\dag}}{{\rm Tr}(K_{n}\rho K_{n}^{\dag})})
\frac{\mathcal{M}_{w}(\rho)K_{n}\widetilde{\tau}K_{n}^{\dag}}{{\rm Tr}(K_{n}\rho K_{n}^{\dag}){\rm Tr}(\frac{\mathcal{M}_{w}(\rho)K_{n}\widetilde{\tau}K_{n}^{\dag}}{{\rm Tr}(K_{n}\rho K_{n}^{\dag})})})\\
&=\frac{\mathcal{M}_{w}(\rho){\rm Tr}(K_{n}\widetilde{\tau}K_{n}^{\dag})}{{\rm Tr}(K_{n}\rho K_{n}^{\dag})}\mathcal{M}_{w}(\frac{K_{n}\widetilde{\tau}K_{n}^{\dag}}{{\rm Tr}(K_{n}\widetilde{\tau}K_{n}^{\dag})})\\
&\leq\frac{\mathcal{M}_{w}(\rho){\rm Tr}(K_{n}\widetilde{\tau}K_{n}^{\dag})}{{\rm Tr}(K_{n}\rho K_{n}^{\dag})}.
\end{array}
\end{equation}\\
In Eq.(\ref{q}),  the first inequality  is based on that $\mathcal{M}_{w}(\rho)$ satisfies  (S4); the second inequality comes from the fact $0\leq\mathcal{M}_{w}(\frac{K_{n}\widetilde{\tau}K_{n}^{\dag}}{{\rm Tr}(K_{n}\widetilde{\tau}K_{n}^{\dag})})\leq 1$; the second equality hods because  $\frac {K_{n}\widetilde{\delta}K_{n}^{\dag}}{{\rm Tr}(K_{n}\widetilde{\delta}K_{n}^{\dag})}$ is a free state.

So, one gets
\begin{equation}
\begin{aligned}
{\rm Tr}(K_{n}\rho K_{n}^{\dag})\mathcal{M}_{w}(\frac{K_{n}\rho K_{n}^{\dag}}{{\rm Tr}(K_{n}\rho K_{n}^{\dag})})\leq \mathcal{M}_{w}(\rho){\rm Tr}(K_{n}\widetilde{\tau}K_{n}^{\dag}).
\end{aligned}
\end{equation}
By summing  both sides over $n$, we have
\begin{equation}\label{3}
\begin{array}{ll}
\sum\limits_{n}{\rm Tr}(K_{n}\rho K_{n}^{\dag})\mathcal{M}_{w}(\frac{K_{n}\rho K_{n}^{\dag}}{{\rm Tr}(K_{n}\rho K_{n}^{\dag})})
&\leq\sum\limits_{n}\mathcal{M}_{w}(\rho){\rm Tr}(K_{n}\widetilde{\tau}K_{n}^{\dag})
=\mathcal{M}_{w}(\rho)\sum\limits_{n}{\rm Tr}(K_{n}\widetilde{\tau}K_{n}^{\dag})\\
&=\mathcal{M}_{w}(\rho){\rm Tr}(\sum\limits_{n}K_{n}\widetilde{\tau}K_{n}^{\dag})\\
&=\mathcal{M}_{w}(\rho){\rm Tr}(\widetilde{\tau})\\
&=\mathcal{M}_{w}(\rho).
\end{array}
\end{equation}
Here the result that the free operation is trace-preserving has been used.

Finally, we prove $\mathcal{M}_{w}(\rho)$ satisfies (S2). Let $\Lambda$ be a free operation, then
\begin{equation}\label{g}
\begin{array}{ll}
\mathcal{M}_{w}(\Lambda(\rho))=\mathcal{M}_{w}(\sum\limits_{n}{\rm Tr}(K_{n}\rho K_{n}^{\dag})\frac{K_{n}\rho K_{n}^{\dag}}{{\rm Tr}[K_{n}\rho K_{n}^{\dag}]})\leq\sum\limits_{n}{\rm Tr}(K_{n}\rho K_{n}^{\dag})\mathcal{M}_{w}(\frac{K_{n}\rho K_{n}^{\dag}}{{\rm Tr}[K_{n}\rho K_{n}^{\dag}]})\leq \mathcal{M}_{w}(\rho).
\end{array}
\end{equation}
In Eq.(\ref{g}), the first  inequality takes the fact that $\mathcal{M}_{w}(\rho)$ satisfies (S4), the second  inequality   is true, since  $\mathcal{M}_{w}(\rho)$ satisfies (S3). Eq.(\ref{g}) implies that $\mathcal{M}_{w}(\rho)$ meets  (S2).

Thus we have proved the above theorem.

\begin{corollary}\label{1}{\rm There is an upper bound $\mathcal{C}_{d}\mathcal{M}_{w}(\rho)$ for any superposition measure $\mathcal{C}(\rho)$. Here $\mathcal{C}_{d}$ is the value of $\mathcal{C}(\rho)$ for the maximum superposition state, and $\mathcal{M}_{w}(\rho)$ stands for the superposition measure based on weight.}
\end{corollary}

\textit{\rm \textbf{Proof.}} Suppose free state $\widetilde{\delta}$ and resource state $\widetilde{\tau}$ are the optimal states for $\rho$ to minimize $\mathcal{M}_{w}(\rho)$. So $\rho=[1-\mathcal{M}_{w}(\rho)]\widetilde{\delta}+\mathcal{M}_{w}(\rho)\widetilde{\tau}$. Because of the convexity of $C(\rho)$, then
\begin{equation}\label{3}
\begin{array}{ll}
\mathcal{C}(\rho)&\leq[1-\mathcal{M}_{w}(\rho)]\mathcal{C}(\widetilde{\delta})+\mathcal{M}_{w}(\rho)\mathcal{C}(\widetilde{\tau})
=\mathcal{M}_{w}(\rho)\mathcal{C}(\widetilde{\tau})
\leq \mathcal{M}_{w}(\rho)\mathcal{C}_{d}.
\end{array}
\end{equation}
So Corollary 4 is true.

\section{The generalization of superposition resource theory}\label{sec:review}

In this section we will generalize the superposition resource theory from the following two perspectives.

\subsection{Convex function }\label{sec:review}

 Assume that $\mathcal{F}$ is a subset of quantum state space, and  there is a convex function $\mathcal{M}$ which  maps every state in the set $\mathcal{F}$  to zero and maps the other state to a positive real number. The operation $\Lambda$ constituted by Kraus operators $\{K_{n}\}$ is called free operation if it satisfies the condition
\begin{equation}
\begin{aligned}
\sum\limits_{n}{\rm Tr}(K_{n}\rho K_{n}^{\dag})\mathcal{M}(\frac{K_{n}\rho K_{n}^{\dag}}{{\rm Tr}(K_{n}\rho K_{n}^{\dag})})\leq \mathcal{M}(\rho)
\end{aligned}
\end{equation}
for every quantum state $\rho$. Then one has the following conclusion.

\begin{theorem}\label{2}{\rm The convex function $\mathcal{M}$ is the superposition measure with free operation set $\{\Lambda\}$ \cite{46}.}
\end{theorem}

\textit{\rm \textbf{Proof.}} Let's first prove that free operation $\Lambda$  maps a free state to a free state. Since $\mathcal{M}$ is a convex function, then
\begin{equation}
\begin{aligned}
\mathcal{M}(\Lambda(\rho))\leq\sum\limits_{n}{\rm Tr}(K_{n}\rho K_{n}^{\dag})\mathcal{M}(\frac{K_{n}\rho K_{n}^{\dag}}{{\rm Tr}(K_{n}\rho K_{n}^{\dag})})\leq \mathcal{M}(\rho).
\end{aligned}
\end{equation}
If $\rho$ is free state, then $\mathcal{M}(\rho)=0$.  Thus $\mathcal{M}(\Lambda(\rho))=0$. It implies that $\Lambda(\rho)$ is also a free state.

In the case where $\{\Lambda\}$ is a free operation set, obviously the convex function $\mathcal{M}$ satisfies (S1), (S2), (S3), (S4), so $\mathcal{M}$ is a valid measure with free operation set  $\{\Lambda\}$.

\textit{\em \textbf{Example 2.}} Suppose $\{|i\rangle\}_{i=1}^{d}$ is an orthonormal basis and $V$ is a full rank $d\times d$ real matrix. We define $|c_{i}\rangle=V|i\rangle$. For any state $\rho=\sum\limits_{i,j}\rho_{ij}|c_{i}\rangle\langle c_{j}|$ if every $\rho_{ij}$ is real, one defines the state $\rho$ as free state \cite{33,34}. It is well known that there is a convex function $\mathcal{M}(\rho)=S(\rho\|\Delta(\rho))$, where $\Delta(\rho)=\frac{1}{2}(\rho+\rho^{\prime})$, \noindent$\rho^{\prime}=\sum\limits_{ij}\rho_{ji}|c_{i}\rangle\langle c_{j}|$. Next we will show that the Kraus operators of the free operation $\Lambda$ are $\{K_{n}=\sum\limits_{ij}c_{ij}^{n}|c_{i}^{\bot}\rangle\langle c_{j}^{\bot}|\}$, where $c_{ij}^{n}$ is a real number for any $i$, $j$, $n$, and $\sum\limits_{n}K_{n}^{\dag}K_{n}=\mathbb{I}$.  Furthermore we will also demonstrate that the convex function $\mathcal{M}(\rho)$ is a superposition measure.

First we show that $S(\rho\|\Delta(\rho))=0$ holds only for  the state $\rho$ whose $\rho_{ij}$ are real number. Let us  show that $\Delta(\rho)$ is a trace-preserving map. Note that $|c_{i}\rangle=V_{i1}|1\rangle+V_{i2}|2\rangle+\ldots+V_{id}|d\rangle$, and the matrix elements of $V_{ij}$ are all real numbers for arbitrary $i, j=1,2,\ldots,d$, one obtains the transposition matrix of quantum state $\rho$,
\begin{equation}
\begin{aligned}
\rho^{\intercal}&=(\sum\limits_{ij}\rho_{ij}|c_{i}\rangle\langle c_{j}|)^{\intercal}
=\sum\limits_{ij}\rho_{ij}(|c_{j}\rangle\langle c_{i}|)^{*}
=\sum\limits_{ij}\rho_{ij}|c_{j}\rangle\langle c_{i}|
=\rho^{\prime}.
\end{aligned}
\end{equation}
Therefore, ${\rm Tr}(\Delta(\rho))=\frac {1}{2}({\rm Tr}\rho+{\rm Tr}\rho')=\frac {1}{2}({\rm Tr}\rho+{\rm Tr}\rho^{\intercal})={\rm Tr}\rho$. It means  $\Delta(\rho)$ is a trace-preserving map.  Later on, we will show that $\rho=\Delta(\rho)$ only if $\rho_{ij}$ is real. Evidently, $\rho^{\dag}=\sum\limits_{ij}\rho_{ij}^{*}|c_{j}\rangle\langle c_{i}|=\sum\limits_{ij}\rho_{ji}^{*}|c_{i}\rangle\langle c_{j}|$.  As $\rho^{\dag}=\rho$, so $\rho_{ji}^{*}=\rho_{ij}$. Therefore
\begin{equation}
\begin{aligned}
\rho+\rho^{\prime}&=\sum\limits_{ij}(\rho_{ij}+\rho_{ji})|c_{i}\rangle\langle c_{j}|
=\sum\limits_{ij}(\rho_{ij}+\rho_{ij}^{*})|c_{i}\rangle\langle c_{j}|.
\end{aligned}
\end{equation}
Hence, if $\rho=\Delta(\rho)$, we have $\rho_{ij}=\rho_{ij}^{*}$, that means $\rho_{ij}$ are real. Thus $S(\rho\|\Delta(\rho))$ can map just the state whose $\rho_{ij}$ are real to 0.
Therefore we have demonstrated   that $S(\rho\|\Delta(\rho))=0$ holds only for  the state $\rho$ whose $\rho_{ij}$ are real number.

Next we will show that $S(\rho\|\Delta(\rho))$ is convex. By the joint convexity of relative entropy \cite{47},
\begin{equation}
\begin{aligned}
S(\rho\|\sigma)\leq\sum\limits_{i}p_{i}S(\rho_{i}\|\sigma_{i}),
\end{aligned}
\end{equation}
where $\rho=\sum\limits_{i}p_{i}\rho_{i}$, $\sigma=\sum\limits_{i}p_{i}\sigma_{i}$. So
\begin{equation}
\begin{aligned}
S(\rho\|\Delta(\rho))&=S(\sum\limits_{i}p_{i}|\psi_{i}\rangle\langle\psi_{i}|\|\Delta(\sum\limits_{i}p_{i}|\psi_{i}\rangle\langle\psi_{i}|))
=S(\sum\limits_{i}p_{i}|\psi_{i}\rangle\langle\psi_{i}|\|\sum\limits_{i}p_{i}\Delta(|\psi_{i}\rangle\langle\psi_{i}|))\\
&\leq\sum\limits_{i}p_{i}S(|\psi_{i}\rangle\langle\psi_{i}|\|\Delta(|\psi_{i}\rangle\langle\psi_{i}|)),
\end{aligned}
\end{equation}
where $\rho=\sum\limits_{i}p_{i}|\psi_{i}\rangle\langle\psi_{i}|$. It  implies $S(\rho\|\Delta(\rho))$ is convex.

Finally, we're going to show how to obtain $\{K_{n}\}$. According to Ref.\cite{39}, for any completely positive trace-preserving map formed by Kraus operator $\{L_{n}\}$, we have
\begin{equation}
\begin{aligned}
\sum\limits_{n}{\rm Tr}(L_{n}\rho L_{n}^{\dag})S(\frac{L_{n}\rho L_{n}^{\dag}}{{\rm Tr}(L_{n}\rho L_{n}^{\dag})}\|\frac{L_{n}\delta L_{n}^{\dag}}{{\rm Tr}(L_{n}\delta L_{n}^{\dag})})\leq\sum\limits_{n}S(L_{n}\rho L_{n}^{\dag}\|L_{n}\delta L_{n}^{\dag})\leq S(\rho\|\delta).
\end{aligned}
\end{equation}
Therefore, if the Kraus operators $\{K_{n}\}$ which constitute $\Lambda$ satisfy $K_{n}\Delta(\rho)K_{n}^{\dag}=\Delta(K_{n}\rho K_{n}^{\dag})$, then the following result holds
\begin{equation}
\begin{aligned}
&\quad\sum\limits_{n}{\rm Tr}(K_{n}\rho K_{n}^{\dag})S(\frac{K_{n}\rho K_{n}^{\dag}}{{\rm Tr}(K_{n}\rho K_{n}^{\dag})}\|\Delta(\frac{K_{n}\rho K_{n}^{\dag}}{{\rm Tr}(K_{n}\rho K_{n}^{\dag})}))\\
&=\sum\limits_{n}{\rm Tr}(K_{n}\rho K_{n}^{\dag})S(\frac{K_{n}\rho K_{n}^{\dag}}{{\rm Tr}(K_{n}\rho K_{n}^{\dag})}\|\frac{\Delta(K_{n}\rho K_{n}^{\dag})}{{\rm Tr}(K_{n}\rho K_{n}^{\dag})})\\
&=\sum\limits_{n}{\rm Tr}(K_{n}\rho K_{n}^{\dag})S(\frac{K_{n}\rho K_{n}^{\dag}}{{\rm Tr}(K_{n}\rho K_{n}^{\dag})}\|\frac{\Delta(K_{n}\rho K_{n}^{\dag})}{{\rm Tr}(\Delta(K_{n}\rho K_{n}^{\dag}))})\\
&=\sum\limits_{n}{\rm Tr}(K_{n}\rho K_{n}^{\dag})S(\frac{K_{n}\rho K_{n}^{\dag}}{{\rm Tr}(K_{n}\rho K_{n}^{\dag})}\|\frac{\Delta(K_{n}\rho K_{n}^{\dag})}{{\rm Tr}(K_{n}\Delta(\rho)K_{n}^{\dag})})\\
&=\sum\limits_{n}{\rm Tr}(K_{n}\rho K_{n}^{\dag})S(\frac{K_{n}\rho K_{n}^{\dag}}{{\rm Tr}(K_{n}\rho K_{n}^{\dag})}\|\frac{K_{n}\Delta(\rho) K_{n}^{\dag}}{{\rm Tr}(K_{n}\Delta(\rho)K_{n}^{\dag})})\\
&\leq\sum\limits_{n}S(K_{n}\rho K_{n}^{\dag}\|K_{n}\Delta(\rho)K_{n}^{\dag})\\
&\leq S(\rho\|\Delta(\rho)).
\end{aligned}
\end{equation}

Next we give a concrete form of the Kraus operator $\{K_{n}\}$ of $\Lambda$ satisfying $K_{n}\Delta(\rho)K_{n}^{\dag}=\Delta(K_{n}\rho K_{n}^{\dag})$. Choose $K_{n}=\sum\limits_{ij}c_{ij}^{n}|c_{i}^{\bot}\rangle\langle c_{j}^{\bot}|$. Evidently,
\begin{equation}
\begin{aligned}
\Delta(K_{n}\rho K_{n}^{\dag})&=\Delta(\sum\limits_{ij}c_{ij}^{n}|c_{i}^{\bot}\rangle\langle c_{j}^{\bot}|\sum\limits_{\alpha\beta}\rho_{\alpha\beta}|c_{\alpha}\rangle\langle c_{\beta}|\sum\limits_{ml}(c_{ml}^{n})^{*}|c_{l}^{\bot}\rangle\langle c_{m}^{\bot}|)\\
&=\Delta(\sum\limits_{ijml}c_{ij}^{n}\rho_{jl}(c_{ml}^{n})^{*}|c_{i}^{\bot}\rangle\langle c_{m}^{\bot}|)\\
&=\sum\limits_{im}{\rm Re}(\sum\limits_{jl}c_{ij}^{n}\rho_{jl}(c_{ml}^{n})^{*})|c_{i}^{\bot}\rangle\langle c_{m}^{\bot}|,\\
K_{n}\Delta(\rho)K_{n}^{\dag}&=\sum\limits_{ij}c_{ij}^{n}|c_{i}^{\bot}\rangle\langle c_{j}^{\bot}|\sum\limits_{\alpha\beta}{\rm Re}(\rho_{\alpha\beta})|c_{\alpha}\rangle\langle c_{\beta}|\sum\limits_{ml}(c_{ml}^{n})^{*}|c_{l}^{\bot}\rangle\langle c_{m}^{\bot}|\\
&=\sum\limits_{im}\sum\limits_{jl}c_{ij}^{n}{\rm Re}(\rho_{jl})(c_{ml}^{n})^{*}|c_{i}^{\bot}\rangle\langle c_{m}^{\bot}|.
\end{aligned}
\end{equation}
So when $c_{ij}^{n}$ is a real number for any $i$, $j$, $n$, we have ${\rm Re}(c_{ij}^{n}\rho_{jl}(c_{ml}^{n})^{*})=c_{ij}^{n}{\rm Re}(\rho_{jl})(c_{ml}^{n})^{*}$ for any $j$, $l$. Therefore $\Delta(K_{n}\rho K_{n}^{\dag})=K_{n}\Delta(\rho)K_{n}^{\dag}$.

Thus   when the Kraus operators of the free operation $\{\Lambda\}$ are $\{K_{n}=\sum\limits_{ij}c_{ij}^{n}|c_{i}^{\bot}\rangle\langle c_{j}^{\bot}|\}$, where $c_{ij}^{n}$ is a real number for any $i$, $j$, $n$, and $\sum\limits_{n}K_{n}^{\dag}K_{n}=\mathbb{I}$, the convex function $\mathcal{M}(\rho)=S(\rho\|\Delta(\rho))$ is a superposition measure.

\subsection{Operators }\label{sec:review}

Suppose that  there are operators $\{E_{i}\}_{i=1}^{n}$ acting on the Hibert space, where $E_{i}E_{j}=E_i\delta_{ij}$.  It is easy to see that the following operators
\begin{equation}\label{3}
\begin{array}{ll}
E_{1}=\sum\limits_{i=1}^{d_{1}}|c_{i}\rangle\langle c_{i}^{\bot}|,~
E_{2}=\sum\limits_{i=d_{1}+1}^{d_{2}}|c_{i}\rangle\langle c_{i}^{\bot}|,~
E_{3}=\sum\limits_{i=d_{2}+1}^{d_{3}}|c_{i}\rangle\langle c_{i}^{\bot}|,~
\cdots,~
E_{n}=\sum\limits_{i=d_{n-1}+1}^{d}|c_{i}\rangle\langle c_{i}^{\bot}|,
\end{array}
\end{equation}
satisfy the requirement, where $d>d_{n-1}>d_{n-2}>\cdots>d_{2}>d_{1}\geq1$.

Then we define  $\sigma=\frac{\sum\limits_{i=1}^{n}E_{i}\rho E_{i}^{\dag}}{{\rm Tr}(\sum\limits_{i=1}^{n}E_{i}\rho E_{i}^{\dag})}$ as the free states, where $\rho$ is an arbitrary  state in the Hilbert space.
The set of free states is denoted by $\mathcal{F^{\prime}}$. Let $K_{n}=\sum\limits_{i}E_{f_{n}(i)}c_{n,i}E_{i}$ \cite{48}, where $f_{n}(i)$ is index permutation function and $c_{n,i}$ is complex matrix. Obviously,
\begin{equation}\label{3}
\begin{array}{ll}
K_{n}\sigma K_{n}^{\dag}
&=\sum\limits_{i}E_{f_{n}(i)}c_{n,i}E_{i}\sum\limits_{\alpha=1}E_{\alpha}\rho E_{\alpha}^{\dag}\sum\limits_{j}E_{j}^{\dag}c_{n,j}^{\dag}E_{f_{n}(j)}^{\dag}/{{\rm Tr}(\sum\limits_{i=1}^{n}E_{i}\rho E_{i}^{\dag})}\\
&=\sum\limits_{i}E_{f_{n}(i)}c_{n,i}E_{i}\rho E_{i}^{\dag}c_{n,i}^{\dag}E_{f_{n}(i)}^{\dag}/{{\rm Tr}(\sum\limits_{i=1}^{n}E_{i}\rho E_{i}^{\dag})}.
\end{array}
\end{equation}
 Define the free operation $\Lambda=\{K_n|\sum_nK_nK_n^+=I\}$. Clearly $\Lambda$  maps a free state to a free state. So with the free operation set $\{\Lambda\}$ one has  the following two superposition measures.

(1) Generalized superposition measure based on weight:
\begin{equation}\label{3}
\begin{array}{ll}
\mathcal{M}_{w}(\rho)=\min\limits_{\tau~\rm density ~matrix}\{1-\lambda\geq 0|\rho=\lambda\delta+(1-\lambda)\tau,\delta\in\mathcal{F^{\prime}}\}.
\end{array}
\end{equation}

(2) Generalized superposition measure based on robustness:
\begin{equation}\label{3}
\begin{array}{ll}
\mathcal{M}_{{\rm R}}(\rho)=\min\limits_{\tau~\rm density ~matrix}\{s\geq 0|\frac{\rho+s\tau}{1+s}\in\mathcal{F^{\prime}}\}.
\end{array}
\end{equation}

It is easy to see from the Sec.IV, with the good definition of free states and free operations, generalized superposition measure $\mathcal{M}_{w}(\rho)$ satisfies (S1), (S2), (S3) and (S4). By Ref.\cite{49}, we can obtain that generalized superposition measure $\mathcal{M}_{{\rm R}}(\rho)$ satisfies (S1), (S2), (S3) and (S4) with free states and free operations well defined.

\section{conclusion}\label{sec:conclusion}

In this paper, we present three methods of constructing superposition measure, which are methods based on convex roof extended,  state transformation and  weight, respectively. We demonstrate that $\mathcal{M}_{\rm rel.net}^{\prime}(\rho)$ and $\mathcal{M}_{l_{1}}^{\prime}(\rho)$ are superposition measures, and that $\mathcal{M}_{ l_{1}}^{\prime}(\rho)\geq\mathcal{M}_{ l_{1}}(\rho)$. We prove that $\mathcal{C}_{d}\mathcal{M}_{w}(\rho)$ is an upper bound of any superposition measure $\mathcal{C}(\rho)$, where $\mathcal{C}_{d}$ is the value of $\mathcal{C}(\rho)$ for the maximum superposition state. Our approach provides new insights on better understanding of superposition measures. We also generalize the superposition resource theory with convex function $\mathcal{M}$, and give a method to find free operation and new superposition measure. Finally, we generalize the superposition resource theory based on operators, and obtain two good superposition measures with well-defined free operation. We hope that these works can  help us to better understand the resource theory of quantum superposition.

\begin{acknowledgments}
This work was supported by the National Natural Science Foundation of China under Grant No. 12071110, the Hebei Natural Science Foundation of China under Grant No. A2020205014,  funded by Science and Technology Project of Hebei Education Department under Grant Nos. ZD2020167, ZD2021066, and  the Hebei Central Guidance on Local Science and Technology Development Foundation of China under Grant No. 226Z0901G.

\end{acknowledgments}

%\bibliographystyle{apsrev4-1}
%\bibliography{../Some_Tools/mybib}

\begin{thebibliography}{99}

\bibitem{1} Bennett C H, Brassard G, Cr\'{e}peau C, et al. Teleporting an unknown quantum state via dual classical and Einstein-Podolsky-Rosen channels. Phys Rev Lett, 1993, 70(13): 1895-1899.
\bibitem{2}   Deng F G, Long G L,  Liu X S. Two-step quantum direct communication protocol using the Einstein-Podolsky-Rosen pair block. Phys Rev A, 2003,  68(4): 042317.

\bibitem{3} Yan F L, Zhang X Q. A scheme for secure direct communication using EPR pairs and teleportation. Eur Phys J B, 2004, 41(1): 75-78.
\bibitem{4} Bennett C H, Brassard G. Quantum cryptography: public-key distribution and coin tossing. In: Proceedings of IEEE International Conference on Computers, System and Signal Processing. Bangalore: IEEE, 1984, 175-179.
\bibitem{5} Ekert A K. Quantum cryptography based on Bell's theorem. Phys Rev Lett, 1991, 67(6): 661-663.
\bibitem{6} Bennett C H, Brassard G, Mermin N D. Quantum cryptography without Bell's theorem. Phys Rev Lett, 1992, 68(5): 557-559.
\bibitem{7} Zhou Y H, Yu Z W, Wang X B. Making the decoy-state measurement-device-independent quantum key distribution practically useful. Phys Rev A, 2016, 93(4): 042324.
\bibitem{8} Hillery M, Bu\v{z}ek V, Berthiaume A. Quantum secret sharing. Phys Rev A, 1999, 59(3): 1829-1834.

\bibitem{9} Einstein A, Podolsky B, Rosen N. Can quantum-mechanical description of physical reality be considered complete?. Phys Rev, 1935, 47(10): 777-780.
\bibitem{21} Werner R F. Quantum states with Einstein-Podolsky-Rosen correlations admitting a hidden-variable model. Phys Rev A, 1989, 40(8): 4277-4281.
\bibitem{22} Peres A. Separability criterion for density matrices. Phys Rev Lett, 1996, 77(8): 1413-1415.

\bibitem{Horodecki}  Horodecki R,  Horodecki P, Horodecki  M, et al. Quantum entanglement. Rev Mod Phys, 2009,  81(2):  865-942.
\bibitem{18} Brunner N, Cavalcanti D, Pironio S, et al. Bell nonlocality. Rev Mod Phys, 2014, 86(2): 419-478.

\bibitem{23} Hong Y, Gao T, Yan F L. Measure of multipartite entanglement with computable lower bounds. Phys Rev A, 2012, 86(6): 062323.
\bibitem{24} Gao T, Hong Y, Lu Y, et al. Efficient \emph{k}-separability criteria for mixed multi-partite quantum states. Europhys Lett, 2013, 104(2): 20007.
\bibitem{25} Gao T, Yan F L, van Enk S J. Permutationally invariant part of a density matrix and nonseparability of \emph{N}-qubit states. Phys Rev Lett, 2014, 112(18): 180501.


\bibitem{37} {\AA}berg J. Quantifying superposition. arXiv: quant-ph/0612146.
\bibitem{38} Baumgratz T, Cramer M, Plenio M B. Quantifying coherence. Phys Rev Lett, 2014, 113(14): 140401.
\bibitem{35} Yao Y, Xiao X, Ge L, et al. Quantum coherence in multipartite systems. Phys Rev A, 2015, 92(2): 022112.

\bibitem{17} Streltsov A, Adesso G, Plenio M B. Colloquium: quantum coherence as a resource. Rev Mod Phys, 2017, 89(4): 041003.


\bibitem{36} Streltsov A, Rana S, Boes P, et al. Structure of the resource theory of quantum coherence. Phys Rev Lett, 2017, 119(14): 140402.

\bibitem{26} Goold J, Huber M, Riera A, et al. The role of quantum information in thermodynamics-a topical review. J Phys A: Math Theor, 2016, 49(14): 143001.
\bibitem{27} Gour G, M\"{u}ller M P, Narasimhachar V, et al. The resource theory of informational nonequilibrium in thermodynamics. Phys Rep, 2015, 583: 1-51.
\bibitem{28} Bartlett S D, Rudolph T, Spekkens R W. Reference frames, superselection rules, and quantum information. Rev Mod Phys, 2007, 79(2): 555-609.
\bibitem{29} Gour G, Marvian I, Spekkens R W. Measuring the quality of a quantum reference frame: the relative entropy of frameness. Phys Rev A, 2009, 80(1): 012307.
\bibitem{30} Vaccaro J A, Anselmi F, Wiseman H M, et al. Tradeoff between extractable mechanical work, accessible entanglement, and ability to act as a reference system, under arbitrary superselection rules. Phys Rev A, 2008, 77(3): 032114.
\bibitem{31} Gour G, Spekkens R W. The resource theory of quantum reference frames: manipulations and monotones. New J Phys, 2008, 10(3): 033023.
\bibitem{32} Kristj\'{a}nsson H, Chiribella G, Salek S, et al. Resource theories of communication. New J Phys, 2020, 22(7): 073014.
\bibitem{33} Wu K D, Kondra T V, Rana S, et al. Operational resource theory of imaginarity. Phys Rev Lett, 2021, 126(9): 090401.
\bibitem{34} Wu K D, Kondra T V, Rana S, et al. Resource theory of imaginarity: quantification and state conversion. Phys Rev A, 2021, 103(3): 032401.


\bibitem{10} Brand\~{a}o F G S L, Gour G. Reversible framework for quantum resource theories. Phys Rev Lett, 2015, 115(7): 070503.
\bibitem{11} Brand\~{a}o F G S L, Gour G. Erratum: reversible framework for quantum resource theories [Phys. Rev. Lett. 115, 070503 (2015)]. Phys Rev Lett, 2015, 115(19): 199901.
\bibitem{12} Chitambar E, Gour G. Quantum resource theories. Rev Mod Phys, 2019, 91(2): 025001.
\bibitem{13} Gour G. Quantum resource theories in the single-shot regime. Phys Rev A, 2017, 95(6): 062314.
\bibitem{14} Liu Z W, Hu X Y, Lloyd S. Resource destroying maps. Phys Rev Lett, 2017, 118(6): 060502.
\bibitem{15} Renes J M. Relative submajorization and its use in quantum resource theories. J Math Phys, 2016, 57(12): 122202.
\bibitem{16} Liu C L, Yu X D, Tong D M. Flag additivity in quantum resource theories. Phys Rev A, 2019, 99(4): 042322.

\bibitem{19} Bu K F, Anand N, Singh U. Asymmetry and coherence weight of quantum states. Phys Rev A, 2018, 97(3): 032342.
\bibitem{20} Skrzypczyk P, Navascu\'{e}s M, Cavalcanti D. Quantifying Einstein-Podolsky-Rosen steering. Phys Rev Lett, 2014, 112(18): 180404.



\bibitem{39} Theurer T, Killoran N, Egloff D, et al. Resource theory of superposition. Phys Rev Lett, 2017, 119(23): 230401.
\bibitem{40} Torun G, \c{S}enya\c{s}a H T, Yildiz A. Resource theory of superposition: state transdormations. Phys Rev A, 2021, 103(3): 032416.
\bibitem{41} \c{S}enya\c{s}a H T, Torun G. Golden states in resource theory of superposition.  Phys Rev A, 2022, 105(4): 042410.
\bibitem{42} Killoran N, Steinhoff F E S, Plenio M B, et al. Converting nonclassicality into entanglement. Phys Rev Lett, 2016, 116(08): 080402.
\bibitem{43} Yu D H, Zhang L Q, Yu C S. Quantifying coherence in terms of the pure-state coherence. Phys Rev A, 2020, 101(6): 062114.
\bibitem{44} Zhao M J, Ma T, Pereira R. Average quantum coherence of pure-state decomposition. Phys Rev A, 2021, 103(4): 042428.
\bibitem{45} Yao Y, Li D, Sun C P. Anomalies of the weight-based coherence measure and mixed maximally coherent states. Phys Rev A, 2020, 102(3): 032406.
\bibitem{46} Kim S, Xiong C H, Kumar A, et al. Quantifying dynamical coherence with coherence measures. Phys Rev A, 2021, 104(1): 012404.
\bibitem{47} Nielsen M A, Chuang I L. Quantum Computation and Quantum Information(\emph{10th Anniversary Edition}). Cambridge: Cambridge University Press, 2010.
\bibitem{48} Bischof F, Kampermann H, Bru{\ss} D. Quantifying coherence with respect to general quantum measurement. Phys Rev A, 2021, 103(3): 032429.
\bibitem{49} Piani M, Cianciaruso M, Bromley T R, et al. Robustness of asymmetry and coherence of quantum states. Phys Rev A, 2016, 93(4): 042107.
\end{thebibliography}

\end{document}